\makeatletter
\newcommand{\dontusepackage}[2][]{%
  \@namedef{ver@#2.sty}{9999/12/31}%
  \@namedef{opt@#2.sty}{#1}}
\makeatother
\dontusepackage{subfigure}
\documentclass[journal, 12pt, onecolumn]{IEEEtran}
\usepackage{amssymb,amsmath,amsbsy,bm}
\usepackage{fixltx2e} 
\IfFileExists{upquote.sty}{\usepackage{upquote}}{}
\usepackage[T1]{fontenc}
\usepackage[utf8]{inputenc}
\IfFileExists{microtype.sty}{\usepackage{microtype}}{}
\usepackage{natbib}
\usepackage{listings}
\usepackage{longtable,booktabs}
\usepackage{graphicx}
\makeatletter
\def\ScaleIfNeeded{%
  \ifdim\Gin@nat@width>\linewidth
    \linewidth
  \else
    \Gin@nat@width
  \fi
}
\makeatother
\let\Oldincludegraphics\includegraphics
{
 \catcode`\@=11\relax
 \gdef\includegraphics{\@ifnextchar[{\Oldincludegraphics}{\Oldincludegraphics[width=\ScaleIfNeeded]}}
}
\usepackage{caption}
\usepackage{float}
\usepackage{subfig}
\usepackage{algorithm} 
\let\scholmdAlgorithm\algorithm
\let\endscholmdAlgorithm\endalgorithm
\let\algorithm\relax \let\endalgorithm\relax

{
 \catcode`\*=11\relax
 \global\let\scholmdAlgorithm*\algorithm*
 \global\let\endscholmdAlgorithm*\endalgorithm*
 \global\let\algorithm*\relax 
 \global\let\endalgorithm*\relax
}

\usepackage[unicode=true]{hyperref}
\hypersetup{breaklinks=true,
            bookmarks=true,
            pdfauthor={},
            pdftitle={Ultra-low memory seismic inversion with randomized trace estimation},
            colorlinks=true,
            citecolor=black,
            urlcolor=blue,
            linkcolor=black,
            pdfborder={0 0 0}}
\newcommand{\pluseq}{\mathrel{+}=}

\title{Ultra-low memory seismic inversion with randomized trace estimation}
\author{Mathias Louboutin\textsuperscript{1} and Felix J.
Herrmann\textsuperscript{1,2}\\\textsuperscript{1} School of Earth and
Atmospheric Sciences, Georgia Institute of
Technology\\\textsuperscript{2} School of Computational Science and
Engineering, Georgia Institute of Technology\\}
\date{}

\begin{document}
\maketitle

\section{Summary:}\label{summary}


Inspired by recent work on extended image volumes that lays the ground
for randomized probing of extremely large seismic wavefield matrices, we
present a memory frugal and computationally efficient inversion
methodology that uses techniques from randomized linear algebra. By
means of a carefully selected realistic synthetic example, we
demonstrate that we are capable of achieving competitive inversion
results at a fraction of the memory cost of conventional full-waveform
inversion with limited computational overhead. By exchanging memory for
negligible computational overhead, we open with the presented technology
the door towards the use of low-memory accelerators such as GPUs.


\section{Introduction}\label{introduction}


Wave-equation based seismic inversion has resulted in tremendous
improvements in subsurface imaging over the past decade. However, the
computational cost, and above-all memory cost, of these methods remains
extremely high limiting their widespread adaptation. These memory costs
stem from the requirement of the adjoint-state method
\citep[\citet{tarantola}]{lionsjl1971} to store (in memory, on disk,
possibly compressed) the complete time history of the forward modeled
wavefield prior to on-the-fly crosscorrelation with the time-reversed or
adjoint wavefield. Unfortunately, the need to store these forward
wavefields for each source may require allocation of up to terabytes of
memory during high-frequency 3-D imaging. To tackle this memory
requirement and to open the way to the use of low-memory accelerators
such as GPUs, different methods have been proposed that balance memory
usage with computational overhead to reduce the memory footprint. One of
the earliest methods in this area is optimal checkpointing
\citep[\citet{Symes2007}]{Griewank}, which has recently been extended to
include wavefield compression \citep{kukrejacomp}. Given available
memory, optimal checkpointing reduces storage needs at the expense of
having to recompute the forward wavefield. While the resulting memory
savings can be significant, the computational overhead can be high (up
to tens of extra wave-equations). Alternatively, by relying on
time-reversibility of the (attenuation-free) wave-equation,
\citet{McMechan}, \citet{Mittet}, \citet{RaknesR45} proposed an approach
where the forward wavefield is reconstructed during back propagation
from values stored on the boundary. Even though these approaches have
been applied successfully in practice, their implementation becomes
elaborate especially in situations of complex wave physics such as in
elastic or transverse-tilted anisotropic media.

Aside from the aforementioned methods that are in principle error free,
except perhaps for methods that involve lossy compression, approximate
methods have also been proposed to reduce the memory imprint of gradient
calculations for wave-equation based imaging. The rationale behind these
approximations is that by giving up accuracy one may gain
computationally and/or memorywise. This was, for example, the main idea
behind imaging with phase-encoded sources \citep[\citet{krebs},
\citet{Moghaddamsefwi}, \citet{haberse}, \citet{leeuwenfwi}]{romero}.
According to insights from stochastic optimization, there are strong
arguments that accurate gradients are indeed unnecessary especially at
the beginning of the optimization and as long as the approximate
gradient equals the true gradient in expectation \citep{friedlander}. We
make use of this fundamental observation and propose a new, simple to
implement, alternative formulation that allows for approximate gradient
calculations with a significantly reduced memory imprint. At first
sight, our method may be somewhat reminiscent of memory-reducing methods
based on the Fourier transform, which compute the Fourier transform
either on the fly \citep[\citet{witte2018cls}]{Sirgue2010} or in windows
\citep{Nihei2007}. Our method is based on a randomized trace estimation
technique instead. Like Fourier methods, which derive their advantage
from computing the gradient for a relatively small number of Fourier
modes, the proposed randomized algorithm collects compressed data during
forward propagation. This leads to significant memory savings at a
manageable computational overhead (similar if not cheaper than
checkpointing \citep{witte2019TPDedas}). However, compared to optimal
checkpointing and boundary methods, Fourier techniques involve
inaccurate gradients that may result in coherent artifacts during the
inversion. Contrarily, our randomized linear algebra based method
produces incoherent artifacts, easily handled by regularizations
methods, with a computational overhead of at least half of that of
Fourier methods thanks to real-valued arithmetic.

Inspired by recent work on randomized linear algebra for seismic
inversion \citep[\citet{yang2020lrpo}]{vanleeuwen2015GPWEMVA}, we
propose an alternative method based on random trace estimation. Unlike
direct time subsampling \citep{louboutin2015segtcs}, randomized trace
estimation involves matrix probing with random vectors
\citep[\citet{hutchpp}]{Avron} (vectors with random $\pm 1$'s or
Gaussian vectors) frequently employed by randomized algorithms, such as
the randomized SVD \citep{halkorand}. These probings are computationally
efficient because they only involve actions of the matrix on a small
number of random vectors. This allows us to compress the time axes
during forward propagation, greatly reducing memory usage in a
computationally efficient way, yielding gradient calculations with a
controllable error.

Our contributions are organized as follows. First, we introduce
randomized trace estimation including a recently proposed
orthogonalization step that improves its performance and accuracy. Next,
we show that gradients calculated with the adjoint-state method for
wave-equation based inversion can be approximated by random trace
estimation. Compared to conventional gradient calculations, the
approximated gradient leads to significant memory reductions and faster
computation of certain imaging conditions. To further justify the
proposed algorithm, we discuss how its computational and memory
requirements compare to existing methods. We conclude by illustrating
the advocacy of the proposed method on a 2D synthetic full-waveform
inversion example.


\section{Methodology}\label{methodology}


In this section, we briefly lay out the key components of our seismic
inversion method that requires significantly less memory. We start by
introducing stochastic estimates for the trace of a matrix, followed by
how these estimates can be used to approximate the gradient of a
time-domain formulation of the adjoint state method for the
wave-equation. We conclude by providing estimates of memory use of the
proposed method and how it compares to other memory reducing methods.

\subsection{Randomized trace
estimation}\label{randomized-trace-estimation}

To address memory pressure of data-intensive applications, a new
generation of randomized algorithms have been proposed. Contrary to
classical deterministic techniques that aim for maximal accuracy, these
methods are stochastic in nature and provide answers with a controllable
error. Examples of these techniques are the randomized SVD
\citep[\citet{yang2020lrpo}]{halkorand} and randomized trace estimation
\citep[Avron,][]{hutchpp}. The latter was used to justify wave-equation
inversions with random phase encoding
\citep[\citet{leeuwenfwi}]{haberse}. In this work, we also rely on
randomized trace estimation but now to reduce the memory imprint of
wave-equation based inversion. At its heart, randomized trace estimation
\citep[\citet{hutchpp}]{Avron} derives from the following approximation
of the identity $\mathbf{I}$:
\begin{equation}
\begin{aligned}
\operatorname{tr}(\mathbf{A}) &=\operatorname{tr}\left(\mathbf{A} \mathbb{E}\left[\mathbf{z} \mathbf{z}^{\top}\right]\right) =\mathbb{E}\left[\operatorname{tr}\left(\mathbf{A} \mathbf{z} \mathbf{z}^{\top}\right)\right]\\
&= \mathbb{E}\left[\mathbf{z}^{\top} \mathbf{A} \mathbf{z}\right] \approx \frac{1}{r}\sum_{i=1}^r\left[\mathbf{z}^{\top}_i \mathbf{A} \mathbf{z}_i\right]= \frac{1}{r}\operatorname{tr}\left(\mathbf{Z}^\top \mathbf{A}\mathbf{Z}\right)
\end{aligned}
\label{randomtrace}
\end{equation}
 where the $\mathbf{z}_i$'s are the random probing vectors for which
$\mathbb{E}(\mathbf{z}^{\top}\mathbf{z})=1$ and $\mathbb{E}$ is the
stochastic expectation operator. The above estimator is unbiased (exact
in expectation) and converges to the true trace of the matrix
$\mathbf{A}$, i.e. $\operatorname{tr}(A)=\sum_i \mathbf{A}_{ii}$, with
an error that decays with $r$ and without access to the entries of
$\mathbf{A}$. Only actions of $\mathbf{A}$ on the probing vectors are
needed and we will exploit this property and the factored form of the
matrix $\mathbf{A}$ in gradient calculations for wave-equation based
inversion. Motivated by recent work
\citep[\citet{graff2017SINBADFlrp}]{hutchpp} we will also employ a
partial \texttt{qr} factorization \citep{trefethen1997numerical} that
approximates the range of the matrix $\mathbf{A}$---i.e., we approximate
the trace with probing vectors
$\begin{bmatrix}\mathbf{Q},\thicksim\end{bmatrix} = \operatorname{qr}(\mathbf{A}\mathbf{Z})$
where $\mathbf{Z}$ is a Rademacher random matrix of $\pm 1$.

\subsection{Approximate gradient
calculations}\label{approximate-gradient-calculations}

While one may argue that inaccurate gradients need to be avoided at all
times, approximations to the gradient are actually quite common. For
instance, wave-equation based inversion with phase-encoded sources
derives from a similar approximation but this time on the data misfit
objective. \citet{haberse} showed that this type of approximate
inversion is an instance of random trace estimation. In addition,
possible artifacts can easily be removed by adding regularization,
e.g.~by imposing the TV-norm constraint on the model
\citep[\citet{peters2018pmf}]{esser2016tvr} or by including
curvelet-domain sparsity constraints on Gauss-Newton updates
\citep{li2015GEOPmgn}. Let us consider the standard adjoint-state FWI
problem, which aims to minimize the misfit between recorded field data
and numerically modeled synthetic data \citep[\citet{tarantola},
\citet{virieux}, \citet{louboutin2017fwi},
\citet{louboutin2017fwip2}]{lionsjl1971}. In its simplest form, the data
misfit objective for this problem reads
\begin{equation}
\underset{\mathbf{m}}{\operatorname{minimize}} \ \frac{1}{2} ||\mathbf{F}(\mathbf{m}; \mathbf{q}) - \mathbf{d}_{\text{obs}} ||_2^2
\label{adj}
\end{equation}
 where $\mathbf{m}$ is a vector with the unknown physical model
parameter (squared slowness in the isotropic acoustic case),
$\mathbf{q}$ the sources, $\mathbf{d}_{\text{obs}}$ the observed data
and $\mathbf{F}$ the forward modeling operator. This data misfit is
typically minimized with gradient-based optimization methods such as
gradient descent \citep{plessixasfwi} or Gauss-Newton
\citep{li2015GEOPmgn}. While the presented approach carries over to
arbitrary complex wave physics, we derive our memory reduced gradient
approximation for the isotropic acoustic case where the gradient for a
single source $\delta\mathbf{m}$ can be written as
\begin{equation}
\delta\mathbf{m} = \sum_t \mathbf{\ddot{u}}[t] \mathbf{v}[t]
\label{iccc}
\end{equation}
 where $\mathbf{u}[t], \mathbf{v}[t]$ are the vectorized (along space)
full-space forward and adjoint solutions of the forward and adjoint
wave-equation at time index $t$. The symbol $\ddot{}$ represents
second-order time derivative. To arrive at a form where randomized trace
estimation can be used, we write the above zero-lag crosscorrelation
over time as the trace of the outer product for each space index
$\mathbf{x}$ separately. By using the dot product property,
$\sum \mathbf{x}_i \mathbf{y}_i=\mathbf{x}^\top\mathbf{y}=\operatorname{tr}(\mathbf{x}\mathbf{y}^\top)$,
in combination with Equation~\ref{randomtrace}, we approximate the
gradient via random-trace estimation---i.e.,
\begin{equation}
\begin{split}
\delta\mathbf{m}[\mathbf{x}] = \operatorname{tr}\left(\mathbf{\ddot{u}}[t, \mathbf{x}]\mathbf{v}[t, \mathbf{x}]^\top\right) \approx \frac{1}{r} \operatorname{tr}\left((\mathbf{Q}^\top \mathbf{\ddot{u}}[\mathbf{x}]) (\mathbf{v}[\mathbf{x}]^\top \mathbf{Q})\right) \\
\end{split}
\label{optr}
\end{equation}
 where parenthesis were added to show that the matrix-vector products
between the wavefields and the probing matrix $\mathbf{Q}$ can be
computed independently. We use this property to arrive at our proposed
ultra-low memory unbiased approximation of the gradient summarized in
Algorithm~\ref{pic} below. This algorithm runs for each space index with
the same probing matrix $\mathbf{Q}$. Instead of storing the wavefield
during the forward pass, e.g.~via checkpointing, the sum of the probed
(with $\mathbf{Q}^\top$) wavefield is for each space index accumulated
(line 2) in the variable $\overline{\mathbf{u}}[\mathbf{x}]$. Because of
the probing, we only need to store $N\times r$ with $r\ll n_t$ samples
in $\overline{\mathbf{u}}$ instead of $N\times n_t$ with $n_t$ the
number of timesteps. As we will show, $r$ can be very small compared to
$n_t$, leading to significant memory savings. Similarly, during the back
propagation pass (lines 4-7), we accumulate for each spatial index
$\overline{\mathbf{v}}$. After finishing the back propagation
iterations, the gradient is calculated by computing the trace of the
outer product of the accumulated probed wavefields (line 8). To avoid
forming an unnecessarily large matrix for the outer product in line 8,
we simply compute the sum over the probing size of the pointwise product
of $\overline{\mathbf{u}}$ and $\overline{\mathbf{v}}$ in practice.
Before deriving a practical scheme for FWI based on this probing scheme,
let us first discuss its memory use and how it compares to other known
approaches to reduce the memory imprint of wave-equation based
inversion.

\begin{scholmdAlgorithm}
0.~\textbf{for~t=2:nt-1}~~~~~~~~~~~~\#~forward~propagation\\1.~~~~$\mathbf{u}[t+1] = f(\mathbf{u}[t], \mathbf{u}[t-1], \mathbf{m}, \mathbf{q}[t])$\\2.~~~~$\overline{\mathbf{u}}[\mathbf{x}] \pluseq \mathbf{Q}^\top \mathbf{\ddot{u}}[\cdot, \mathbf{x}]$\\3.~\textbf{end~for}\\4.~\textbf{for~t=nt:-1:1}~~~~~~~~~~~\#~back~propagation\\5.~~~~$\mathbf{v}[t-1] = f^\top(\mathbf{v}[t], \mathbf{v}[t+1], \mathbf{m}, \delta \mathbf{d}[t])$\\6.~~~~$\overline{\mathbf{v}}[\mathbf{x}] \pluseq \mathbf{Q}^\top \mathbf{v}[\cdot, \mathbf{x}]$\\7.~\textbf{end~for}\\8.~output:~$\frac{1}{r}\operatorname{tr}(\overline{\mathbf{u}}\, \overline{\mathbf{v}}^\top)$
\caption{Approximate gradient calculation with random trace
estimation}\label{pic}
\end{scholmdAlgorithm}

\subsection{Memory estimates}\label{memory-estimates}

From Equation~\ref{optr}, we can easily estimate the memory imprint of
our method compared to conventional FWI. For completeness, we also
consider other mainstream low memory methods: optimal checkpointing
\citep[\citet{Symes2007}, \citet{kukrejacomp}]{Griewank}, boundary
methods \citep[\citet{Mittet}, \citet{RaknesR45}]{McMechan}, and DFT
methods \citep[\citet{Sirgue2010}, \citet{witte2018cls}]{Nihei2007}.
This memory overview generalizes to other wave-equations and imaging
conditions easily as our method generalizes to any time-domain
adjoint-state method. We estimate the memory requirements for a
three-dimensional domain with $N=N_x \times N_y \times N_z$ grid points
and $n_t$ time steps. Conventional FWI requires storing the full
time-space forward wavefield to compute the gradient. This requirement
leads to a memory requirement of $N\times n_t$ floating point values.
Our method, on the other hand, for $r$ probing vectors (i.e
$\mathbf{Q} \in \mathbb{R}^{n_t \times r}$), requires only $N\times r$
floating point values during each of the forward and backward passes for
a total of $2\times N\times r$ values. The memory reduction factor is,
therefore, $\frac{n_t}{2 r}$. This memory reduction is similar to
computing the gradient with $\frac{r}{2}$ Fourier modes. We summarize
the memory usage compared to other state-of-the-art algorithms in
table~\ref{memest}.

\begin{table}
\centering
\resizebox{\columnwidth}{!}{\begin{tabular}{cccccc}
\toprule\addlinespace
& FWI & DFT & Probing & Optimal checkpointing & Boundary
reconstruction\tabularnewline
\midrule
Compute & 0 & $\mathcal{O}(2r)\times n_t \times N$ &
$\mathcal{O}(r)\times n_t \times N$ &
$\mathcal{O}(log(n_t))\times N \times n_t$ &
$n_t \times N$\tabularnewline
Memory & $N \times n_t $ & $2r\times N$ & $r \times N$ &
$\mathcal{O}(10) \times N$ & $n_t \times N^{\frac{2}{3}}$\tabularnewline
\bottomrule
\end{tabular}}
\caption{Memory estimates and computational overhead of different
seismic inversion methods for $n_t$ time steps and $N$ grid
points.}\label{memest}
\end{table}

It is worth noting that unlike the other methods in the table, boundary
reconstruction methods tend to have stability issues for more complex
physics, in particular with physical attenuation making it ill-suited
for real-world applications. As stated in the introduction, our method
closely follows the computational and memory cost of Fourier methods by
a factor of two related to real versus complex arithmetic. We also show
that unlike checkpointing or boundary methods, neither the memory nor
computational overhead depends on the number of time steps, therefore
our method offers improved scalability.

\subsection{Imaging conditions}\label{imaging-conditions}

Aside from these clear advantages regarding memory use, the proposed
approximation scheme also has computational advantages when imposing
more elaborate imaging conditions such as the inverse scattering imaging
condition \citep[\citet{witteisic}]{Whitmore} for RTM or wavefield
separation \citep{Faqi} for FWI. In most cases, these imaging condition
can be expressed as linear operators that only act on the spatial
dimensions of the wavefields and not along time. Because these operators
are linear, we can factor these operators out and directly apply them to
the probed wavefields consisting of \texttt{r} reduced time steps rather
than to every time steps with $n_t\gg r$. We can achieve this by making
use of the following identity (the same applies to $\mathbf{v}$):
\begin{equation}
\mathbf{Q}^\top \left(\mathbf{D}_x \mathbf{u}[\cdot,\mathbf{x}]\right) = \mathbf{D}_x \left(\mathbf{Q}^\top \mathbf{u}[\cdot,\mathbf{x}]\right),
\label{commute}
\end{equation}
 which holds as long as the linear imaging condition only acts along the
space directions and not along time. Because $r\ll n_t$ this can lead to
significant computational savings especially in the common situation
where imposing imaging conditions becomes almost as expensive as solving
the wave-equation itself.

\subsection{\texorpdfstring{Choice of the probing matrix
$\mathbf{Q}$}{Choice of the probing matrix \textbackslash{}mathbf\{Q\}}}\label{choice-of-the-probing-matrix-mathbfq}

\begin{figure}
\centering
\includegraphics[width=0.750\hsize]{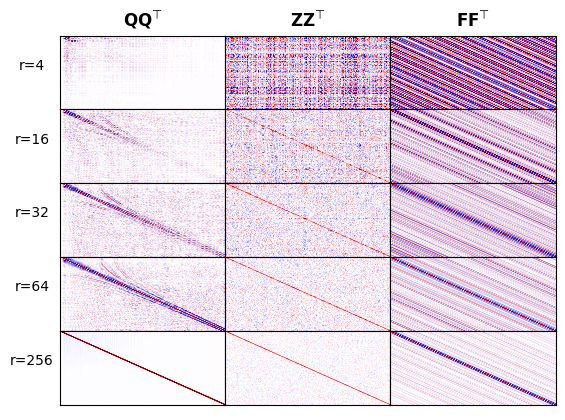}
\caption{Probing vector for varying probing size.}\label{pvec}
\end{figure}

While strictly random, e.g.~random $\pm 1$ as in Rademacher or Gaussians
\citep{Avron} an extra orthogonalization step (via a \texttt{qr}
factorization on random probings $\mathbf{A}\mathbf{Z}$) allows us to
capture the range of $\mathbf{A}$
\citep[\citet{graff2017SINBADFlrp}]{hutchpp}, which leads faster decay
of the error as a function of $r$. This error in the gradient is due to
``cross talk''---i.e. $\mathbf{Z}\mathbf{Z}^\top\ne\mathbf{I}$.
Unfortunately, we do not have easy access to $\mathbf{A}$ during the
approximate gradient calculations outlined in Algorithm~\ref{pic}.
Moreover, orthogonalizing each spatial gridpoint separately would be
computationally infeasible. Despite these complications, we argue that
we can still get a reasonable approximation of the range by random
probing the observed data organized as a matrix for each source
experiment---i.e., we have
\begin{equation}
\begin{bmatrix}\mathbf{Q},\thicksim\end{bmatrix} = \operatorname{qr}(\mathbf{A}\mathbf{Z}) \quad\text{with}\quad \mathbf{A}=\mathbf{D}_{\text{obs}}\mathbf{D}_{\text{obs}}^\top
\label{QR}
\end{equation}
 where the observed data vector $\mathbf{d}_{\text{obs}}$ for each shot
is shaped into a matrix along the time and lumped together receiver
coordinates. Since observed data contains information on the temporal
characteristics of the wavefields, we argue that this outer product can
serve as a proxy for the time characteristics of the wavefield
everywhere. In Figure~\ref{pvec}, we demonstrate the benefits of the
additional orthogonalization step by comparing the outer products of the
Rademacher probing matrix $\mathbf{Z}$, the restricted Fourier matrix
$\mathbf{F}$, and orthogonalized probing vectors $\mathbf{Q}$ as a
function of increasing $r$. The following observations can be made.
First, as expected the ``cross-talk'', i.e.~amplitudes away from the
diagonal, becomes smaller when $r$ increases, which is to be expected
for all cases. However, we observe also that the outer product converges
faster to the identity for both the Rademacher and \texttt{qr} factored
case while coherent artifact remain with Fourier due to the truncation.
Second, because of the orthogonalization the artifacts for the outer
product of $\mathbf{Q}$ are much smaller and this should improve the
accuracy of the gradient at the expense of a relatively minor cost of
carrying our a \texttt{qr} factorization for each shot record. Finally,
compared to the Fourier basis, in combination with source spectrum
informed frequency sampling, our probing factors are informed by
estimates of the range of the sample covariance kernel spanned by the
traces in each shot record.

\section{Example}\label{example}

\begin{figure*}
\centering
\includegraphics[width=1.000\hsize]{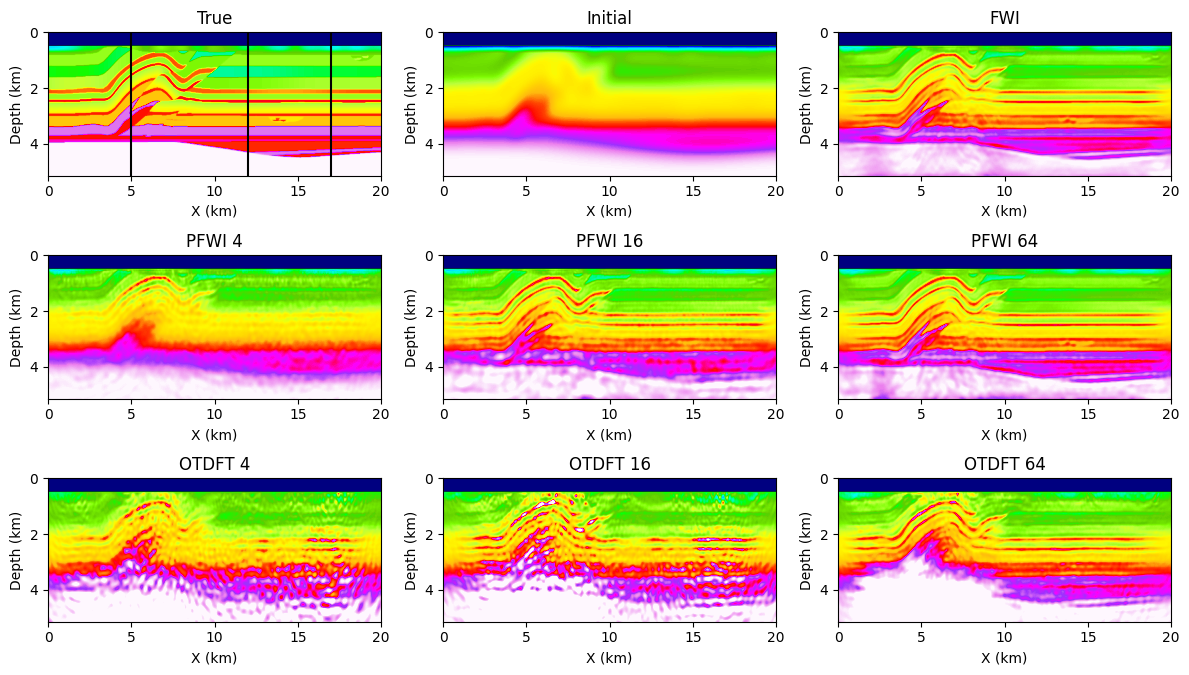}
\caption{Probed versus On-the-fly DFT FWI on the 2D overthrust model
with equivalent memory imprints.}\label{fig:2d-fwi-comp}
\end{figure*}

We illustrate our method on the 2D overthrust model and compare our
inversion results to conventional FWI and on-the-fly DFT
\citep[\citet{witte2018cls}]{Sirgue2010}. We consider a $20$km by $5$km
2D slice of the well-known overthrust model. The dataset consists of
$97$ sources $200$m apart at $50$m depth. Each shot contains between
$127$ and $241$ ocean bottom nodes $50$m apart at $500$m depth for a
maximum offset of $6$km. The data is modeled with an $8$Hz Ricker
wavelet and $3$sec recording. We show the true model and the initial
background model with the inversion results in
Figure~\ref{fig:2d-fwi-comp}. We ran $20$ iterations of Spectral
Projected Gradient (SPG, gradient descent with box constraints
\citep{schmidt09a}) with $20$ randomly selected shots per iteration
\citep{Aravkin11TRridr} in all cases. We can clearly see from
Figure~\ref{fig:2d-fwi-comp} that our probed gradient allows the
inversion to carry towards a good velocity estimate. As theoretically
expected, for few probing vectors, we do not converge since our
approximation is not accurate enough. However, we start to obtain a
result comparable to the true model with as few as $16$ probing vectors.
Additionally, this result could easily be improved by adding constraints
as regularization \citep[\citet{peters2018pmf}]{esser2016tvr}. On the
other hand, we can also see in Figure~\ref{fig:2d-fwi-comp} that for an
equivalent memory cost, on-the-fly DFT fails to converge to an
acceptable result for any number of frequencies, most likely due to the
coherent artifacts that stem from the DFT. These results could also be
improved with constraints or with a better choice of selected
frequencies. We finally compare the three vertical traces highlighted in
black to detail the accuracy of the inverted velocity plotted in
Figure~\ref{fig:2d-fwi-comp}. These traces show that our probed
inversion result is in the vicinity of results obtained with standard
FWI, which itself is close to the true model.

\begin{figure*}
\centering
\captionsetup[subfigure]{labelformat=empty}
\subfloat[]{\includegraphics[width=0.330\hsize]{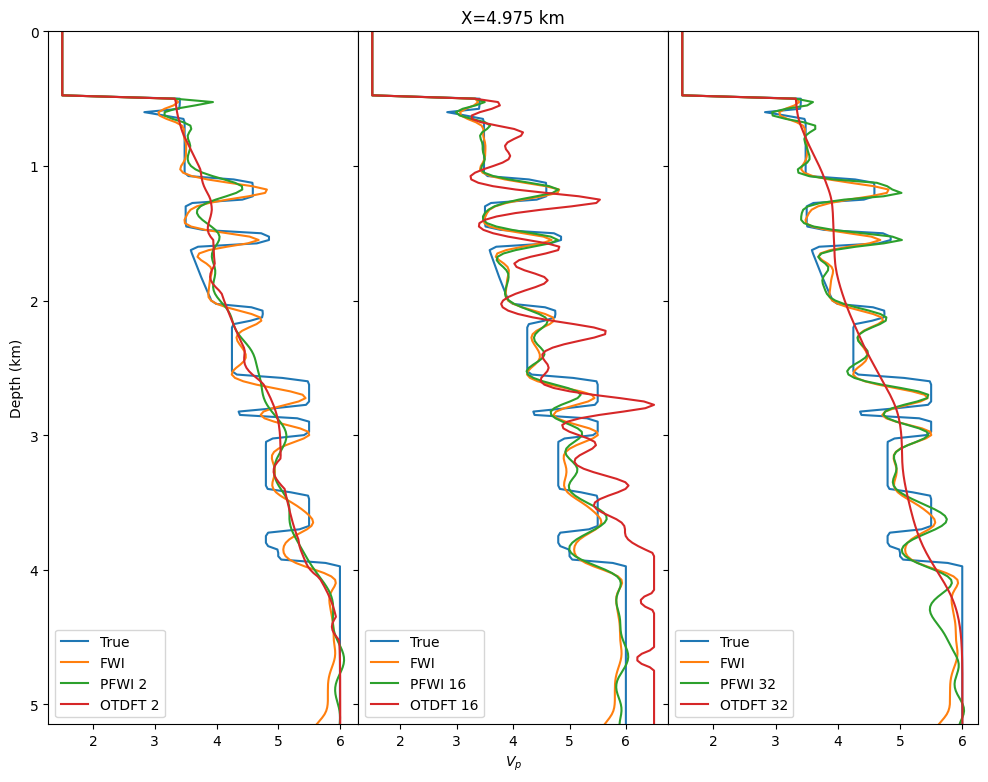}}
\subfloat[]{\includegraphics[width=0.330\hsize]{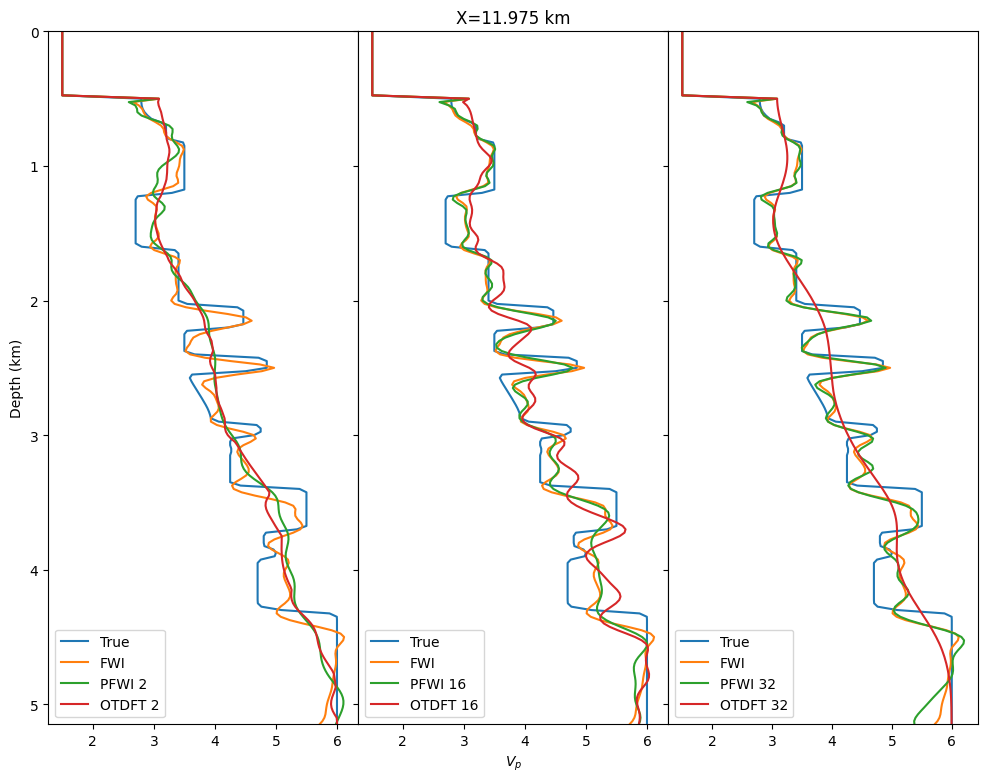}}
\subfloat[]{\includegraphics[width=0.330\hsize]{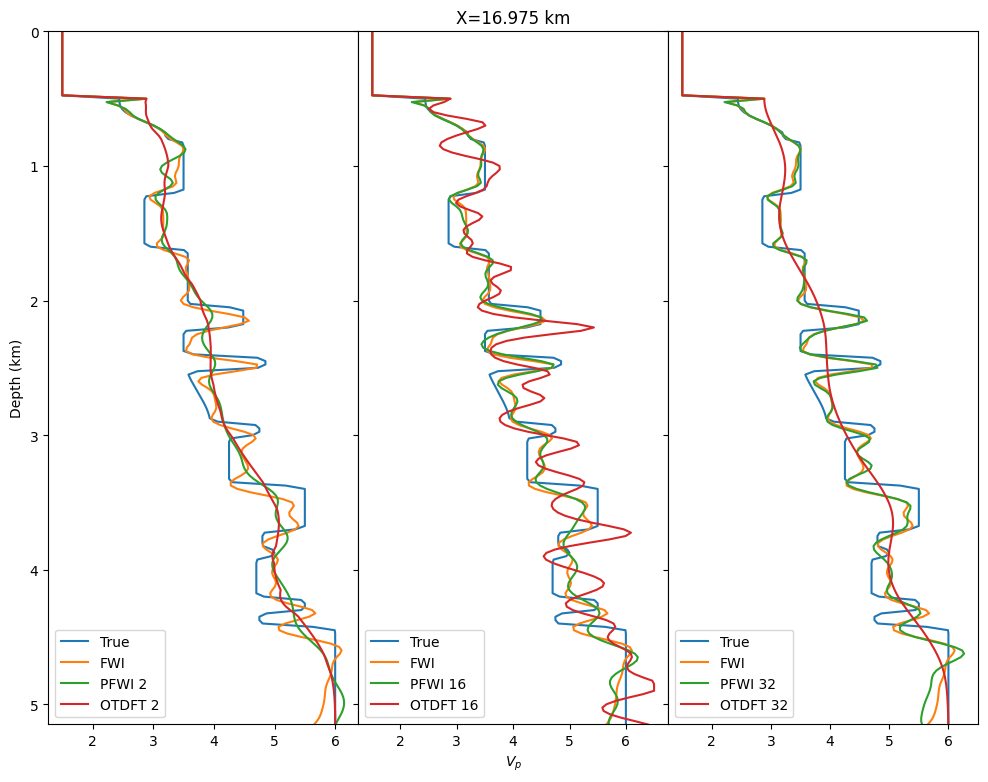}}
\caption{Vertical trace comparison between the OTDFT and probed
FWI.}\label{fig:2d-fwi-tr}
\end{figure*}


\section{Discussion and Conclusions}\label{discussion-and-conclusions}


We introduced a randomized trace estimation technique to drastically
reduce the memory footprint of wave-equation based inversion. We
achieved this result at a computational overhead similar smaller than
that of Fourier-based methods. However, compared to these methods our
approach is simpler and produces less coherent crosstalk. Aside from the
elegance and simplicity of the well-established technique of randomized
trace estimation our proposed approach derives its performance from
probing vectors that approximate the range of the data's sample
covariance operator. We successfully demonstrated our randomized scheme
on a realistic 2D full-waveform example where our method outperforms
Fourier based methods. In future work, we plan to extend our methodology
to high-frequency inversion and to per offset or even per trace probing
vectors instead of per shot. We will also test this method on more
complex wave physics. Our implementation and examples are open sourced
at
\href{https://github.com/slimgroup/TimeProbeSeismic.jl}{TimeProbeSeismic.jl}
(https://github.com/slimgroup/TimeProbeSeismic.jl) and extends our Julia inversion framework
\href{https://github.com/slimgroup/JUDI.jl}{JUDI.jl}. Our code is also
designed to generalize to 3D and more complicated physics as supported
by
\href{https://www.devitoproject.org}{Devito}~\citep[\citet{devito-compiler}]{devito-api}.


\section{Acknowledgement}\label{acknowledgement}


This research was carried out with the support of Georgia Research
Alliance and partners of the ML4Seismic Center.


\bibliography{bibliography}

\end{document}